\documentclass{article}                            
\usepackage{amsmath,amssymb,enumerate,eucal,supertabular}

\def\bn{\bigskip\noindent}
\def\mn{\medskip\noindent}

\def\sn{\smallskip\noindent}

\title{Nemesis encounters of nearby Hipparcos stars}
\author{Igor Yu. Potemine\footnote{Universit\'e 
Paul Sabatier, Institut de Math\'ematiques de Toulouse (IMT), 118 route de 
Narbonne, F-31062 Toulouse Cedex 9, France, 
igor.potemine@math.univ-toulouse.fr}}
\date{}

\begin{document}
\maketitle

\abstract{Very close encounters of stars might lead to significant 
perturbations of their Oort-type clouds, planetesimal belts and planetary 
systems. We have calculated encounter parameters of Hipparcos stars using 
HIP2, Pulkovo and CRVAD2 catalogs of radial velocities. It turns out that 
some stars have encounters within 0.1 pc from each other and might be 
on an essential collision course up to few thousands AU. We present here 
examples with accurate astrometric data and stable with respect to errors 
of radial velocities within $\pm 0.3$ km/s. They include $\beta$ vs. $\gamma$ 
Virginis, 61 Cygni vs. $\chi^1$ Orionis as well as close encounters 
involving $\eta$ Bo\"otis, AB Doradus, 61 Urs{\ae} Majoris and others.}

\section{Introduction}

Thanks to Hipparcos mission quite accurate astrometric data are available 
for many nearby stars within 25 parsecs from the Sun. Main errors in 
kinematical calculations for these stars are now coming from uncertainties 
in their radial velocities. In order to handle this problem, we calculate 
average minimal distances $d_{\mathrm{min}}$ between two stars varying radial 
velocities in the range of $\pm 0.3$ km/s from their catalog values $v_r$ 
(cf.~\cite{G}). More exactly, we calculate the average value of 
$d_{\mathrm{min}}$ with respect to radial velocities $v_r\pm k/10$, 
$0\leq k \leq 3$, for each considered star. It excludes all unstable cases 
when encounters are too distant in time etc. Multiple stars with uncertain 
astrometic data for some massive components are also excluded. Finally, 
alternative values of radial velocities from CRVAD2 catalog are used 
in order to confirm our results.

\section{$\beta$ and $\gamma$ Virginis}

These are two famous bright F-type stars in the constellation Virgo 
at $\approx$ 10.93 and $\approx$ 11.68 parsecs from the Sun resp. 
Gamma Virginis is actually a physical binary of twin F0V-stars. 

\newpage
\tablehead{\hline}\tabletail{\hline}
\begin{supertabular}{|c|c|c|c|c|}
Id &Plx &pmRA &pmDE &RV\\
&(mas) &(mas/yr) &(mas/yr) &(km/s)\\

\hline $\beta$ Vir &91.50 &740.23 &-270.43 &+4.3 / +4.11 \\
\hline $\gamma$ Vir &85.58 &-614.76 &61.34 &-19.9 / -19.56 \\ 
\end{supertabular}

\mn The current distance between $\beta$ and $\gamma$ Virginis is about 2.7 
parsecs. If radial velocities from Pulkovo and CRVAD2 catalogs are accurate 
enough these stars should have the closest encounter within 1/10 parsecs 
from each other. In the table below we express minimal distances in 
astronomical units.

\medskip
\tablehead{\hline}\tabletail{\hline}
\begin{supertabular}{|c|c|c|c|c|}
Id1 &Id2 &Catalog &time (yr) &min. dist. (AU)\\

\hline $\beta$ Vir &$\gamma$ Vir &Pulkovo &+33030 &19660 \\
\hline & &CRVAD2 &+33110 &16360 \\
\end{supertabular}

\section{61 Cygni vs. $\chi^1$ Orionis} 

61 Cygni is 13th nearest star system according to RECONS at about
3.5 parsecs from the Sun. It is a visual binary consisting of two
K-type components HIP 104214 and HIP 104217. 61 Cygni A is slightly 
more massive than its companion. The total proper motion of this 
system exceeds 5000 mas/yr. 

\sn $\chi^1$ Orionis = HIP 27913 is a G0V main-sequence star 
with a faint M-type companion located at $\approx$ 8.66 pc from the Sun.
The current distance between 61 Cygni and $\chi^1$ Orionis is about 10.2
parsecs.

\medskip
\tablehead{\hline}\tabletail{\hline}
\begin{supertabular}{|c|c|c|c|c|}
Id &Plx &pmRA &pmDE &RV\\
&(mas) &(mas/yr) &(mas/yr) &(km/s)\\

\hline 61 Cyg A &286.82 &4168.31 &3269.20 &-65.9 / -66.46 \\
       61 Cyg B &285.88 &4106.90 &3144.68 &-64.3 / -64.9\\
\hline $\chi^1$ Ori &115.43 &-162.54 &-99.51 &-13.1 / -13.45 \\
\end{supertabular}

\bn The average time and the average minimal distance of encounters are 
calculated separately for two components of 61 Cygni using both Pulkovo and 
CRVAD2 catalogs of radial velocities. It follows that the barycenter of 
61 Cygni AB system should have the closest encounter with $\chi^1$ Ori 
within 20,000 AU from each other.

\medskip
\tablehead{\hline}\tabletail{\hline}
\begin{supertabular}{|c|c|c|c|c|}
Id1 &Id2 &Catalog &time (yr) &dist. (AU)\\

\hline $\chi^1$ Ori &61 Cyg A &Pulkovo &+81700 &9880 \\
       &61 Cyg B & &+83420 &21750 \\
\hline
\hline $\chi^1$ Ori &61 Cyg A &CRVAD2 &+81260 &14330 \\
       &61 Cyg B & &+82950 &22750 \\
\end{supertabular}

\section{Other notable encounters}

We present here some interesting examples for stars currently located 
within 25 parsecs of the Sun. They involve some notorious stars 
like a subgiant $\eta$ Bo\"otis. AB Doradus is a rotationally 
variable star belonging to the moving group of the same name.

\medskip
\tablehead{\hline}\tabletail{\hline}
\begin{supertabular}{|c|c|c|c|c|}
Id &Plx &pmRA &pmDE &RV\\
&(mas) &(mas/yr) &(mas/yr) &(km/s)\\

\hline GJ 588 &168.66 &-1117.12 &-1028.54 &+15.5 / +15.4 \\
       GJ 688 &90.91 &-179.30 &-96.60 &+20.0 / +20.46\\

\hline 61 UMa &104.04 &-12.55 &-380.75 &-5.5 / -5.82 \\
       GJ 536 &99.72 &-823.47 &598.19 &-25.8 / -25.8 \\

\hline LHS 2713 &90.32 &632.60 &-262.21 &+7.6 / +5.9 \\
       $\eta$ Boo &87.75 &-60.95 &-356.29 &+0.7 / +0.54\\

\hline AB Dor &65.93 &33.16 &150.83 &+32.4 / +32.7 \\
       GJ 156 &64.24 &-4.95 &529.59 &+62.4 / +61.8 \\
\end{supertabular}

\bn In the table below we indicate the current distance between stars
as well as average times and minimal distances of their close encounters.

\bigskip
\tablehead{\hline}\tabletail{\hline}
\begin{supertabular}{|c|c|c|c|c|}
Id1 &Id2 &distance (pc) &time (yr) &min. dist. (AU)\\

\hline GJ 588 &GJ 688 &8.85 &-159280 &17930 \\
       & & &-158130 &16180 \\
\hline 61 UMa &GJ 536 &8.22 &+122610 &22280 \\
       & & &+122390 &19080 \\      
\hline LHS 2713 &$\eta$ Boo &1.81 &+48780 &26040 \\
       & & &+49140 &11910 \\ 
\hline AB Dor &GJ 156 &15.58 &-206380 &17960 \\
       & & &-206850 &16560 \\
\end{supertabular}

\section{Transit of Luyten 726-8 within 1 ly of Epsilon Eridani}

Though Luyten 726-8 (=BL/UV Ceti) is not a Hipparcos star we have checked 
our previous result about its close encounter with Epsilon Eridani 
(\emph{cf.}~\cite{P2}).

\medskip
\tablehead{\hline}\tabletail{\hline}
\begin{supertabular}{|c|c|c|c|c|}
Id &Plx &pmRA &pmDE &RV\\
&(mas) &(mas/yr) &(mas/yr) &(km/s)\\

\hline BL/UV Ceti &373.70 &3296.00 &563.00 &+21.9 / +29.0 \\
\hline $\varepsilon$ Eri &310.94 &-975.17 &19.49 &+16.3 / +16.15 \\ 
\end{supertabular}

\mn Radial velocities for BL/UV Ceti are taken from Malaroda+ and GCRV catalogs.

\medskip
\tablehead{\hline}\tabletail{\hline}
\begin{supertabular}{|c|c|c|c|c|}
Id1 &Id2 &Catalog &time (yr) &min. dist. (AU)\\

\hline $\varepsilon$ Eri &BL/UV Ceti &Malaroda+ &+31720 &59060 \\
\hline & &GCRV &+31200 &60100 \\
\end{supertabular}

\section{Acknowledgements}

I am very grateful to Stepan Orevkov who has written a compact C++
program calculating encounter parameters of Hipparcos stars from 
their catalog values.

\smallskip This research has made use of the SIMBAD database, operated at 
CDS, Strasbourg, France.

\nocite{*}
\bibliographystyle{plain}
\bibliography{closehip}

\begin{thebibliography}{1}

\bibitem{D}
J.-M. Deltorn and P.~Kalas.
\newblock {\em Search for Nemesis encounters with Vega, $\varepsilon$ Eridani,
  and Fomalhaut}, volume 244 of {\em ASP Conference Series}, pages 227--232.
\newblock San Fransisco: Astronomical Society of the Pacific, 2001.

\bibitem{G}
G.~Gontcharov.
\newblock Pulkovo compilation of radial velocities for 35495
  $\textrm{H}$ipparcos stars.
\newblock {\em AL}, 32(11):759--771, 2006.
\newblock \emph{cf.}~catalogue III/252 at {SIMBAD}.

\bibitem{P3}
I.~Potemine.
\newblock Derived catalog of close encounters of {H}ipparcos stars with
  {P}ulkovo radial velocities.
\newblock \emph{in preparation}.

\bibitem{P1}
I.~Potemine.
\newblock Giant $\textrm{N}$emesis candidate $\textrm{HD}$ 107914 /
  $\textrm{HIP}$ 60503 for the perforation of $\textrm{O}$ort cloud.
\newblock {\em arXiv:1003.5308}, pages 1--3, 2010.

\bibitem{P2}
I.~Potemine.
\newblock Transit of {L}uyten 726-8 within 1 ly from {E}psilon {E}ridani.
\newblock {\em arXiv:1004.1557}, pages 1--3, 2010.

\bibitem{vL}
F.~van Leeuwen.
\newblock Validation of the new $\textrm{H}$ipparcos reduction.
\newblock {\em A{\&}A}, 474:653--664, 2007.
\newblock \emph{cf.}~catalogue I/311 at {SIMBAD}.

\end{thebibliography}

\end{document}